# Heat-capacity anomalies in the presence of high magnetic fields in the spin-chain compound, $Ca_3Co_2O_6$

E.V. Sampathkumaran[a,b,*], Z. Hiroi[a], S. Rayaprol[b] and Y. Uwatoko[a]

[a]*The Institute for Solid State Physics, The University of Tokyo, 5-1-5 Kashiwanoha, Kashiwa, Chiba 277 8581, Japan*

[b]*Tata Institute of Fundamental Research, Homi Bhabha Road, Mumbai 400005, India.*

## Abstract

Heat-capacity behavior in the presence of externally applied magnetic field (H) up to 140 kOe is investigated (2-35 K) for the quasi-1-dimensional compound, $Ca_3Co_2O_6$, believed to order magnetically with the 'so-called' 'partially disordered anti-ferromagnetic structure'. The results reveal that there is a shift of the magnetic transition occurring around 24 K to a marginally *higher* temperature (by 0.5 K) as H is increased from 10 to 30 kOe, uncharacteristic of anti-ferromagnets. No peak attributable to magnetic ordering could be observed in the entire field range of investigation in the vicinity of the second magnetic transition, occurring around 10 K. However, hysteresis studies of H-dependence of C indicate a subtle change in the magnetic structure as the T is lowered across 10 K.

[*]**Corresponding author:** E.V. Sampathkumaran, Tata Institute of Fundamental Research, Homi Bhabha Road, Mumbai-400005, India. Tel: +91-22-2280-4545; Fax: +91 22 2280 4610. E-mail: sampath@tifr.res.in

The compounds with quasi-one-dimensional magnetic behavior have been attracting a lot of attention in recent years. In this respect, the family of compounds of the type $(Sr,Ca)_3ABO_6$ (A, B= a metallic ion), forming in the $K_4CdCl_6$-derived rhombohedral structure, is of interest, as this structure consists of chains formed by face-sharing $AO_6$ trigonal prisms and $BO_6$ octahedra with A/B ions forming a triangular lattice in the basal plane [1]. Among these, the compound, $Ca_3Co_2O_6$, is of special significance as brought out in many recent publications and has generated theoretical interests as well [1-14]: For instance, this is the only compound in which both A and B sites can be fully occupied by the same ion. Two magnetic transitions, one around 24 K and the other around 10 K, have been proposed; the former arises from the anti-ferromagnetic ordering of the ferromagnetic chains at the vortices of the hexagon, with the third chain within triangle in the basal plane remaining disordered. In this sense, judged by the behavior in the temperature range 10-24 K, this compound serves [3] as an example for an uncommon magnetic structure, viz., 'partially disordered anti-ferromagnetic structure (PDA)', similar to the one seen in $CsCoCl_3$. However, the nature of the transition around 10 K remains elusive and the proposals to explain the magnetic state below 10 K can be broadly classified into two categories: (i) ferri-magnetic ordering involving the third chain and (ii) spin-glass freezing of the third chain. Our heat-capacity (C) measurements (in the absence of external magnetic field (H)) revealed [9] a well-defined peak close to 24 K, which could be associated with anti-ferromagnetic ordering. However, no prominent peak could be observed around 10 K which prompted us [9] to propose that the magnetic transition around 10 K is not of a long range type (implying a disordered magnetism); a subsequent report [11] also confirmed this finding. Viewing together all these results, it appears that the understanding of the magnetism of this compound is still open and therefore it is of interest to gather more experimental results on this compound.

In this article, we report the results of our investigation of H dependence of C on this compound, performed up to a magnetic field as high as 140 kOe *for the first time for any compound within family,* with the intention of clarifying (i) whether the magnetic ordering around 24 K could be suppressed by the application of high fields, considering that this transition is essentially anti-ferromagnetic in nature, (ii) whether a peak around 10 K could be seen if there is any H-induced magnetic ordering, and (iii) possible differences in the nature of the magnetic state below and above 10 K, as inferred by C. It may be remarked that the C measurements on this interesting family of oxides is scarce, barring few reports [9, 11, 15-17]. While this work was going on, we saw the report of Hardy et al [11] on the C behavior of this compound in the presence of 0, 20 and 50 kOe; we compare our results with their findings.

Two batches of $Ca_3Co_2O_6$ in the polycrystalline form were prepared as described in Ref. 9 and characterized by X-ray diffraction. The specimens were further characterized by magnetization measurements down to 1.7 K to make sure that the magnetic behavior of the specimens agrees with those reported in the literature. The C measurements as a function of temperature (2 – 35 K) have been carried out employing a commercial Physical Property Measurement System, PPMS, (Quantum Design) in the presence of fixed external fields (0, 10, 30, 50, 100, and 140 kOe) in the persistent mode; for sample 2, the measurements were performed in closer temperature intervals, however restricted to three typical fields (0, 30 and 100 kOe), mainly to verify the trend observed in sample 1 in the field-dependence of the 'first' transition point (that is occurring near 24 K). In addition, the measurements for sample 1 as a function of H at three selected temperatures (5, 12, and 24 K) were also performed.

Heat-capacity behavior as a function of temperature taken in the presence of several magnetic fields is shown in figures 1 and 2 for samples 1 and 2 respectively. The results obtained in zero-field are in good agreement with those from a homemade calorimeter [9], that is, there is a peak around 24.2 K, with a monotonous decrease of C with decreasing temperature. Qualitatively, the features observed in the data are the same for both the samples, though there is a marginal sample dependence of absolute values. As the H is increased up to 30 kOe, the peak height around 24 K grows dramatically and further increase of H diminishes the peak value gradually. We have also verified this in a different way by taking C data as a function of H at 24 K and it is clear from the figure 3 that the C peaks for 30 kOe. The following findings are in agreement with the observations by Hardy et al [11]. The C values in the intermediate temperature range (around 12 K) are the highest for H= 0 and it appears that the application of H modifies the way the magnetic entropy evolves below 24 K. This is demonstrated in the plots of C/T and the total entropy (S), shown in figure 2b and 2c for sample 2, derived from the data measured in 0, 30 and 100 kOe. The S values for H= 0 and 30 kOe at 24 K have been found to be the same. The plot of C/T versus $T^2$ is found to be linear below about 6 K, however yielding a lower value (about 1 mJ mol$^{-1}$K$^{-1}$) for the linear term compared to that in Ref. 11 (10 mJ mol$^{-1}$K$^{-1}$) and a marginally higher value for the Debye temperature (about 440 K instead of 415 K).

The new findings of importance inferred from our data are:

(i) While the temperature at which the peak in C appears is constant (24.2 K) for H= 0 and 10 kOe, there is an upward shift of the curves (peaking at 24.8 K) as the H is increased to 30 kOe (see Fig. 1, inset) without any further change in the peak temperature for a further increase of H. This shift, though small, is of significance, considering that the application of H is actually expected to suppress anti-ferromagnetic ordering. This experimental observation is confirmed in sample 2 as well (see Fig. 2d) by taking data in closer temperature interval. We have verified this on a third sample as well (not shown here). At this juncture, it is important to note that Lashley et al [18] brought out the limitations of commercial PPMS machine to yield reliable C behavior under certain circumstances. We have ensured that the small upward shift of the peak in C is not an experimental artifact (like possible small error in H-dependence of temperature sensor calibration), by performing various standard test runs, for instance, by measuring C of a Gd-based antiferromagnetic metal for which the Neel temperature is around 21 K (for which we have previous C(H) data from a home-made calorimeter). Another point in support of this is the observation of a sudden shift of the peak position when one changes H from 10 to 30 kOe only and this would not be the case if there is a calibration error of the temperature-axis, which is otherwise expected to result in a gradual shift of the peak with H. According to Ref. 18, the relaxation method employed in PPMS instrument can suppress the peak height, say, for a first-order transition; however, such problems are ruled out in our case, as the peak height in C(T) plot for H= 0 is comparable to that obtained by a semi-adiabatic method in a home-made calorimeter [9].

It may be recalled [3-6,9] that there are steps in the isothermal magnetization curves in the magnetically ordered state, and for a field of 10 to 30 kOe, ferri-magnetism involving the 'incoherent chain' develops (above 10 K range), and at much higher fields, all the chains are ferro-magnetically aligned. The present observation therefore implies that the 'incoherent chain' responding to the application of H completely modifies the way the exchange coupling among the chains takes place in the vicinity of first magnetic transition. However, it is interesting that the corresponding increase in the transition temperature due to the realignment of magnetic moments among the chains to result in ferromagnetic structure at higher fields - even as high as 140 kOe – is not significant. This is an unique observation, reported for the first time for this material.

(ii) The prominent peak around 24 K is observable even at the highest field of 140 kOe, whereas in Ref. 11 (which reports C behavior in the presence of relatively low values of H only, viz., at 20 and 50 kOe), the peak could not be observed even at 50 kOe and therefore there is a sudden evolution of magnetic entropy (see a change in the slope of S(T) curve around 25 K for H= 100 kOe in Fig. 2c) at the transition temperature even at very high fields.

(iii) While there is no prominent peak attributable to long range ordering (we mean, a well-defined magnetic structure) around 10 K in zero field, no peak could be observed at any of the fields in that vicinity of temperature. A careful look at the C/T plot (Fig. 2b) reveals that the broad feature appearing in the zero field data around 8-15 K is in fact suppressed at higher fields. It is very difficult to understand this observation with the ideas of ferri-magnetism *exclusively (vide infra),* either in zero-field or in high fields.

The question, one is tempted to ask from the point (iii) made above, is whether there is any magnetic transition at all, around 10 K. In order to address this issue, we have performed C measurements at selected temperatures for up and down field variations (Fig. 3). We find that, at 12 and 24 K, the curves obtained while decreasing H follows the 'up' cycle, whereas, one observes hysteresis effect at 5 K. This suggests that there is a subtle change in the magnetism of this compound as the temperature is lowered across 10 K. At this juncture, we would like to state that there is a microscopic evidence from $^{59}$Co NMR relaxation rates behavior as well [19] for this conclusion.

We now make a proposal for the nature of the 'subtle magnetic phase' at low temperatures (below 10 K). We believe that the solution lies in a combined look at all the available relevant literature, in particular taking into consideration the observation of ferri-magnetism in neutron studies [5] *even in zero field* and possible crystallographic defects. We believe that the magnetic chains need not be of infinite length and can be broken into segments (nano-magnets?) by defects in the chains, say, by the presence of a small fraction of Ca ions along the magnetic chain, which is actually known [20] for some compounds of this family; therefore, it is possible that some segments are ferri-magnetic (which is detected in neutron studies), while the others undergo spin-glass freezing (as detected in other macroscopic measurements like C and magnetization); alternatively all the segments are ferri-magnetic, but they couple with each other in a random manner like a cluster spin-glass (or like a super-paramagnet which could be consistent with a large frequency-dependence of ac susceptibility [6,9] as well). Therefore, the 'subtle' phase below 10 K could be an inhomogeneous magnetic phase of this kind within the homogeneous compound. It should also be noted that the neutron data below 12 K [5] reveal a suppression of ferri-magnetic component by the application of high magnetic fields, which implies destruction of the anti-ferromagnetic alignment of the 'third chain', thereby presumably favoring magnetic frustration. The absence of C anomaly at high fields around 10 K may also originate from this frustration. We hope these observations will be useful for a better understanding of this novel material.

To conclude, heat-capacity measurements up to as high as 140 kOe have been performed on the exotic spin-chain material, $Ca_3Co_2O_6$. The results reveal that there is a finite upward shift of the 'first' (that is, around 24 K) magnetic transition temperature for H > 10 kOe, in contrast to the behavior of anti-ferromagnets, thereby establishing that this compound is an unconventional anti-ferromagnet. No other peak due to long range magnetic ordering (that is, with a well-defined magnetic structure), could be observed even at fields as high as 140 kOe at lower temperatures.

**REFERENCES**

1. H. Fjellvag, E. Gulbrandsen, A. Assland, A. Olsen and B.C. Hauback, J. Solid State Chem. 124, 190 (1996).

2. S. Aasland, H. Fjellvag and B. Hauback, Solid State Commun. 101, 187 (1997).

3. H. Kageyama, K. Yoshimura, K. Kosuge, H. Mitamura and T. Goto, J. Phys. Soc. Japan. 66, 1607 (1997).

4. H. Kageyama, S. Kawasaki, K. Mibu, M. Takano, K. Yoshimura, and K. Kosuge, Phys. Rev. Lett. 79, 3258 (1997).

5. H. Kageyama, K. Yoshimura, K. Kosuge, X. Xu and S. Kawano, J. Phys. Soc. Japan, 67, 357 (1998).

6. A. Maignan, C. Michel, A.C. Masset, C. Martin and B. Raveau, Eur. Phys. J. B 15, 657 (2000).

7. B. Martinez, V. Laukhin, M. Hernando, J. Fontcuberta, M. Parras and J.M. Gonzalez-Calbet, Phys. Rev. B 64, 012417 (2001).

8. B. Raquet, M.N. Baibich, J.M. Broto, H. Rakoto, S. Lambert and A. Maignan, Phys. Rev. B 65, 104442 (2002).

9. S. Rayaprol, K. Sengupta and E.V. Sampathkumaran, Solid State Commun. 128, 79 (2003); arXiv:cond-mat/0306425

10. S. Rayaprol, K. Sengupta and E.V. Sampathkumaran, Proc. Indian. Acad. Sci. (Chem. Sci), 115, 553 (2003).

11. V. Hardy, S. Lambert, M.R. Lees and D. McK Paul, Phys. Rev. B 68, 014424 (2003); arXiv:cond-mat/0307429.

12. R. Vidya, P. Ravindran, H. Fjellvag, A. Kjekshus and O. Eriksson, Phys. Rev. Lett. 91, 186404 (2003).

13. R. Fresard, C. Laschinger, T. Kopp and V. Eyert, Phys. Rev. B 69, 140405(R) (2004).

14. V. Eyert, C. Laschinger, T. Kopp and R. Fresard, Chem. Phys. Lett. 385, 249 (2004).

15. A. Niazi, E.V. Sampathkumaran, P.L. Paulose. D. Eckert, A. Handstein and K.H. Muller, Phys. Rev. B 65, 0644418 (2002).

16. S. Rayaprol, K. Sengupta and E.V. Sampathkumaran, Phys. Rev. B 67, 180404(R) 2003.

17. K. Sengupta, S. Rayaprol, K.K. Iyer and E.V. Sampathkumaran, Phys. Rev. B 68, 012411 (2003).

18. J.C. Lashley et al, Cryogenics 43, 369 (2003).

19. E.V. Sampathkumaran, N. Fujiwara, S. Rayaprol, P.K. Madhu and Y. Uwatoko, Phys. Rev. B, in press; arXiv:cond-mat/0405642.

20. R.C. Layland and H.-C. Loye, J. Alloys and Compd. 299, 118 (2000); K.E. Stitzer, W.H. Henley, J.B. Claridge, H.-C. Loye, and R.C. Layland, J. Solid State Chem. 164, 220 (2002).

Figure 1:

Heat-capacity (C) as a function of temperature in zero-field as well as in the presence of external magnetic fields (H) for the sample 1 of $Ca_3Co_2O_6$. For H= 0 alone, we show the data points to enable the reader to infer the interval of points for all other fields; for other curves, we show only the lines omitting the data points for the sake of clarity (This comment is applicable to figure 2a, 2b and 2c also). Thin vertical lines are drawn through peaks in the inset to show that the peak has shifted as H is increased from 10 to 30 kOe.

Figure 2:

Heat-capacity (C) as a function of temperature in zero-field as well as in the presence of external magnetic fields (H) for the sample 2 of $Ca_3Co_2O_6$, plotted as (a) C vs T and (b) C/T vs T. The total entropy is plotted in (c) and the C data around 24 K is shown in an expanded form in (d). The lines through the data points serve as a guide to the eyes. The curves for H= 30 and 100 kOe overlap below about 20 K in (a), (b) and (c).

Figure 3:

Heat-capacity as a function of magnetic field for $Ca_3Co_2O_6$ at 5, 12 and 24 K, recorded while increasing as well as decreasing the field. The curves for 'up' and 'down' cycles follow each other at 12 and 24 K, whereas at 5 K, there is a hysteresis.

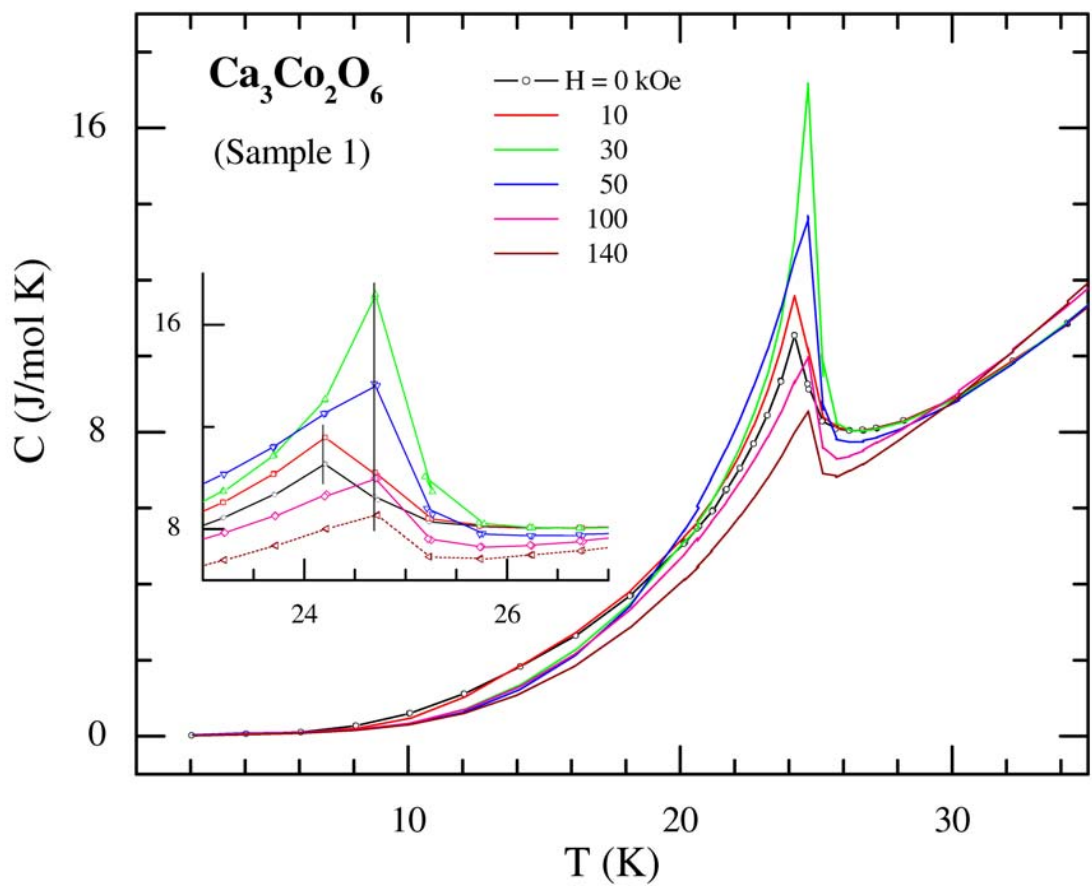

Figure 1

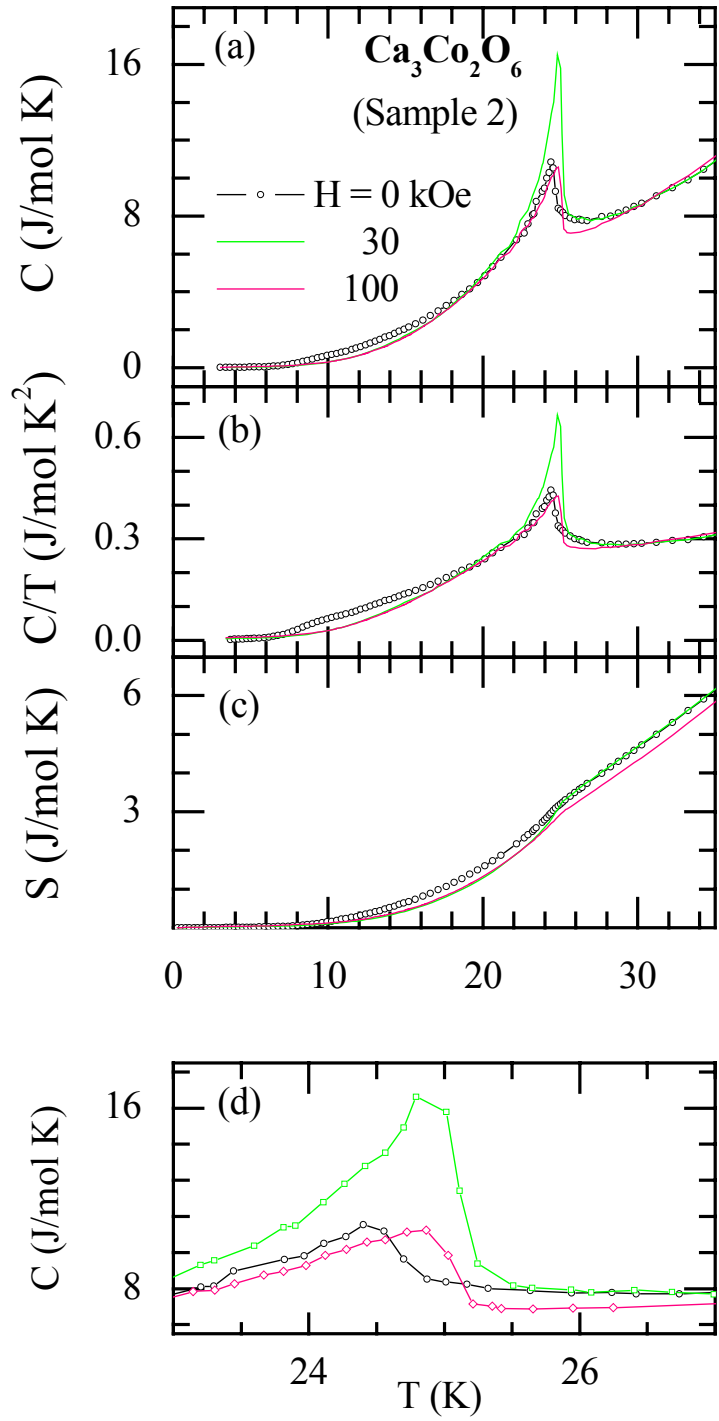

Figure 2

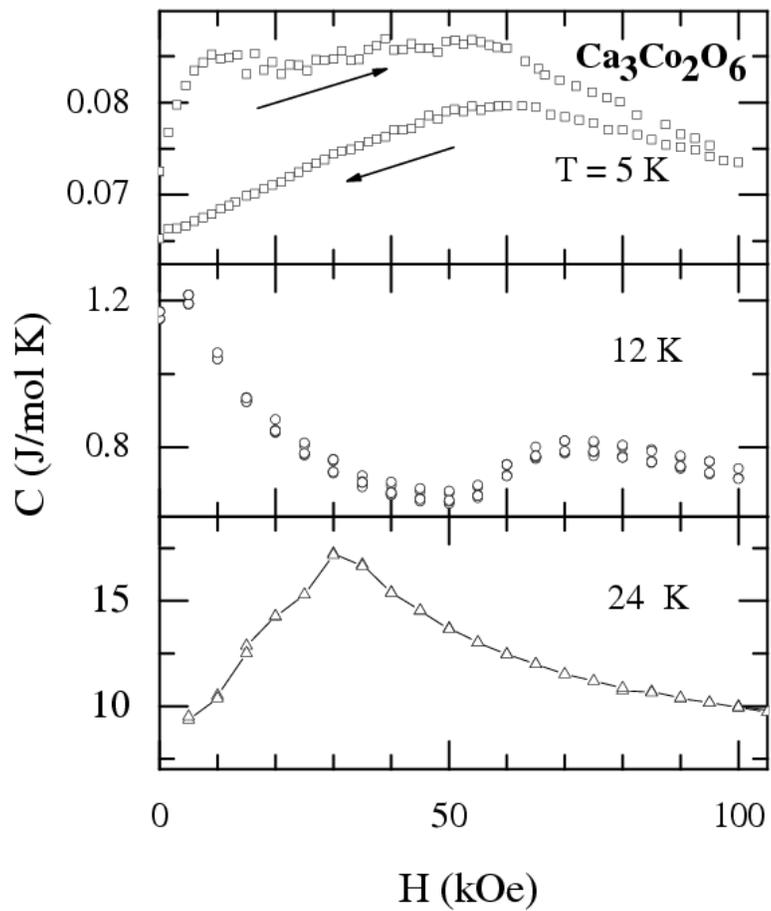

Figure 3